\documentclass[aps,prl,reprint,superscriptaddress]{revtex4-1}
\usepackage[pdftex,bookmarks,colorlinks]{hyperref}
\pdfoutput=1
\usepackage[pdftex]{graphicx}
\usepackage{natbib}
\usepackage{amssymb,amsmath}

\begin{document}



\title{Building micro-soccer-balls with evaporating colloidal fakir drops}



\author{\'Alvaro G. Mar\'in}
\email[]{a.marin@unibw.de}
\affiliation{Physics of Fluids Group, Faculty of Science and Technology, Mesa+ Institute, University of Twente, Enschede, The Netherlands.}

\author{Arturo Susarrey-Arce}
\affiliation{Catalytic Processes and Materials, Faculty of Science and Technology, University of Twente.}
\affiliation{Mesoscale Chemical Systems, Faculty of Science and Technology\\
University of Twente}

\author{Hanneke Gelderblom}
\affiliation{Physics of Fluids Group, Faculty of Science and Technology, Mesa+ Institute, University of Twente, Enschede, The Netherlands.}

\author{Arie van Houselt}
\author{Leon Lefferts}
\affiliation{Catalytic Processes and Materials, Faculty of Science and Technology, University of Twente.}

\author{Han Gardeniers}
\affiliation{Mesoscale Chemical Systems, Faculty of Science and Technology\\
University of Twente}

\author{Detlef Lohse}
\affiliation{Physics of Fluids Group, Faculty of Science and Technology, Mesa+ Institute, University of Twente, Enschede, The Netherlands.}
\author{Jacco H. Snoeijer}
\affiliation{Physics of Fluids Group, Faculty of Science and Technology, Mesa+ Institute, University of Twente, Enschede, The Netherlands.}




\date{\today}

\begin{abstract}
Evaporation-driven particle self-assembly can be used to generate three-dimensional microstructures. We present a new method to create these colloidal microstructures, in which we can control the amount of particles and their packing fraction. To this end, we evaporate colloidal dispersion droplets on a special type of superhydrophobic micro-structured surface, on which the droplet remains in Cassie-Baxter state during the entire evaporative process. The remainders of the droplet consist of a massive spherical cluster of the microspheres, with diameters ranging from a few tens up to several hundreds of microns. We present scaling arguments to show how the final particle packing fraction of these balls depends on the dynamics of the droplet evaporation.
 \end{abstract}


\maketitle

Evaporation-driven particle self-assembly is an ideal mechanism for constructing micro- and nanostructures at scales where direct manipulation is impossible. For example, in colloidal dispersion droplets with pinned contact lines, evaporation gives rise to the so-called coffee-stain effect \cite{Deegan:1997}: a capillary flow drags the particles towards  the contact line to form a ring-shaped stain. Such a flow not only aggregates the particles, but is also able to organize them in crystalline phases \cite{KStebe2004assembly, Bigioni_Witten, Harris:2007, MarinPRL2011}. Similar mechanisms such as the ``convective assembly'' \cite{Velikov:2002, MengNanoletters} are currently successfully used to produce two-dimensional colloidal crystal films.
To obtain three-dimensional clusters of micro-particles, colloidal dispersion droplets which are suspended in emulsions \cite{Velev:Science, manoharan2003science, colloidosomes} or kept in Leidenfrost levitation \cite{Buckling} are used. With these techniques, new colloidal structures arise from the geometrical constraints during the drying \cite{lauga2004evaporation}. This problem of organization of particles into a spherical topography dates back to the days of the first models of the atom and has been extensively studied by Bausch et al. \cite{bausch2003grain, Bausch:Nature}.
The main drawback of these  three-dimensional assembly techniques, however, is the lack of control on both the amount of particles and the particle arrangement in the remaining structures.
\begin{figure}[h*]
\includegraphics[width=3.4 in]{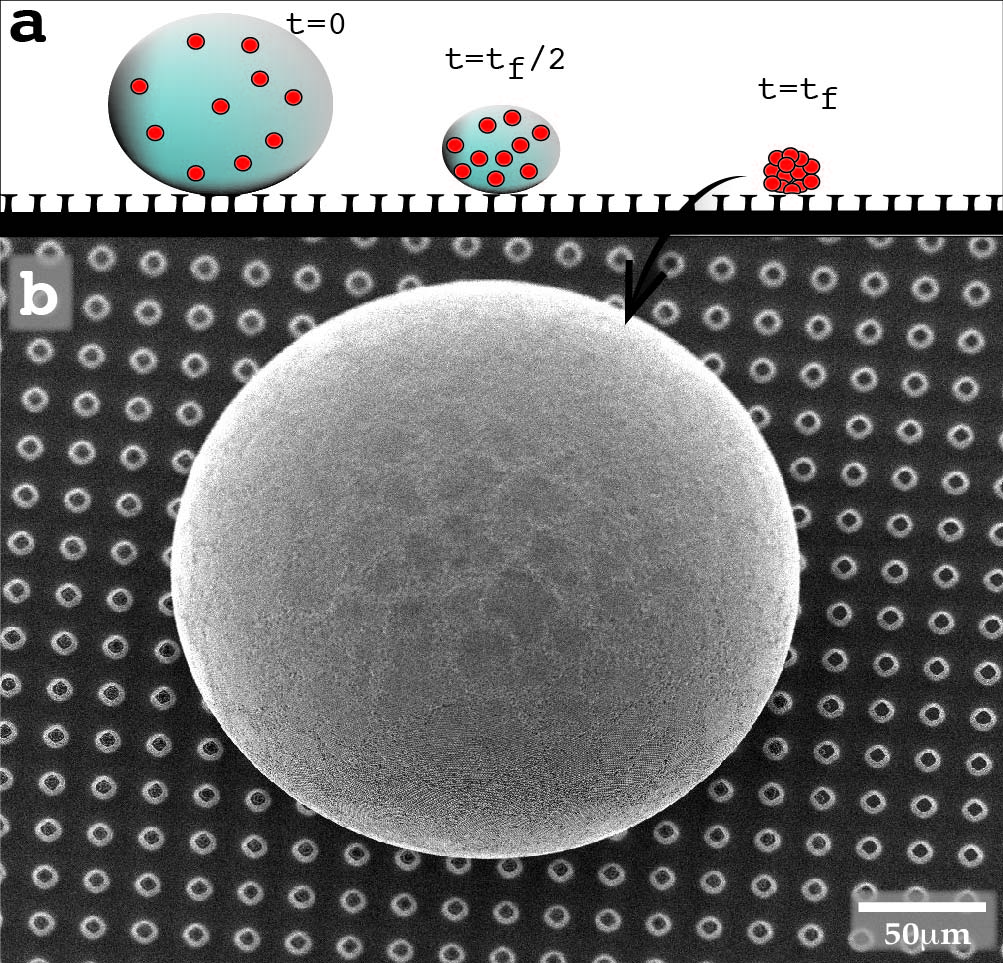}
\caption{(a) A droplet of colloidal solution is left to evaporate on a superhydrophobic surface. As the solvent evaporates, the particle packing fraction increases. Once all the solvent has completely evaporated the colloidal particles have aggregated to form a spherical particle conglomerate: a colloidal supraball.  (b) Top view of the resultant compact colloidal supraball left on the superhydrophobic surface after evaporation. The micropillars forming the structure are seen as circular objects around the supraball.}
\label{fig1}
\end{figure}

\begin{figure}[h]
	\includegraphics[width=3.4 in]{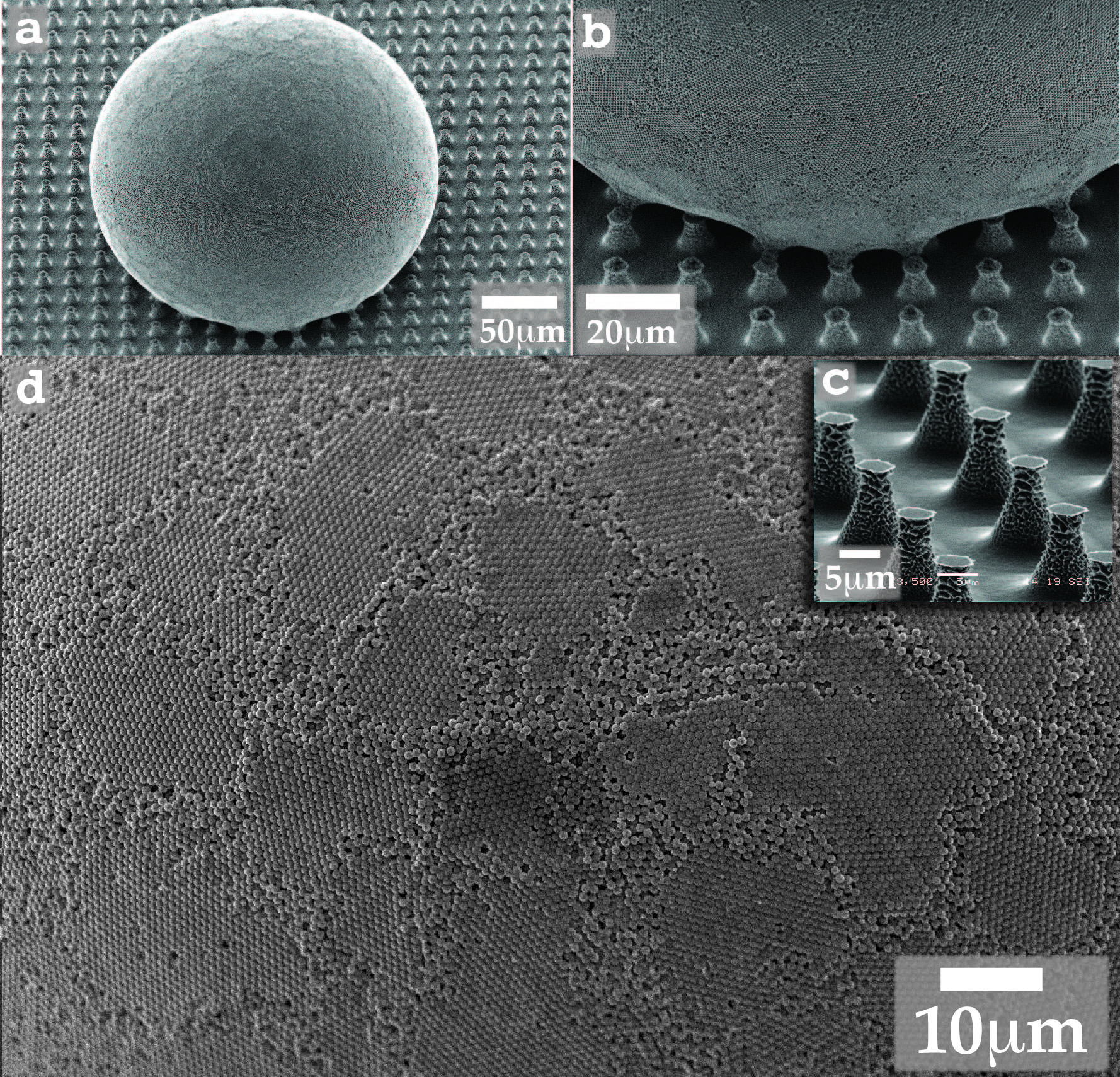}
\caption{(a) Tilted view of the supraball in contact with the microstructure. (b) Detail of the contact area. (c) Magnified view of the micropillars forming the microstructure. (d) Close-up of the supraball surface. The distribution of crystalline patches resemble the pentagons in a soccer-ball.}
\label{fig2}
\end{figure}

In this work, we devise a new, controlled way of generating on-demand self-assembled spherical micro-structures via droplet evaporation on a superhydrophobic surface (see figure \ref{fig1}). We present scaling arguments to predict the particle arrangement in the microstructures formed, based on the dynamics of the evaporation process.
To generate the microstructures, we evaporate colloidal dispersion droplets on a special type of superhydrophobic substrates.
In most of the cases, a liquid Cassie-Baxter state drop evaporating on a superhydrophobic surface will eventually suffer a wetting transition into a Wenzel state, i.e. it will get impaled into the structure and loose its spherical shape \cite{reyssat2008impalement,tsai2010evaporation}. Here, however, we use a surface that combines overhanging pillared structures \cite{tuteja2008} with a hierarchical nano-structure (figure \ref{fig2}c). These surface properties impose a huge energy barrier for the wetting transition to occur, and therefore the droplet will remain almost floating over the structure in a Cassie-Baxter state during its entire life \cite{ArturoPNAS}.
A typical result can be observed in figure \ref{fig1}: a water droplet containing $1\mu m$ polystyrene particles (concentration 0.08\% weight and initial volume $5\mu l$) evaporates on the superhydrophobic surface at room temperature and 30\% humidity. After a typical evaporation time of 45 minutes, the solvent is completely evaporated and only the colloids are left upon the substrate. Remarkably,  the particles ($\sim10^7$ in this particular case) are not just lying scattered over the substrate but they have aggregated and form a spherical macro-cluster resting on top of the micro-pillars, which we call \emph{colloidal supraball}. We do not observe shell formation and buckling during the evaporation of the droplets \cite{Buckling}: the supraballs we obtain are solid, and present a high mechanical resistance and stability. When looking closer at the surface of these particular colloidal supraballs as shown in figure \ref{fig2}d, one can identify crystalline flat patches which resemble the pentagonal patches present in a soccer ball.

\begin{figure}[h*]
	\includegraphics[width=3.4 in]{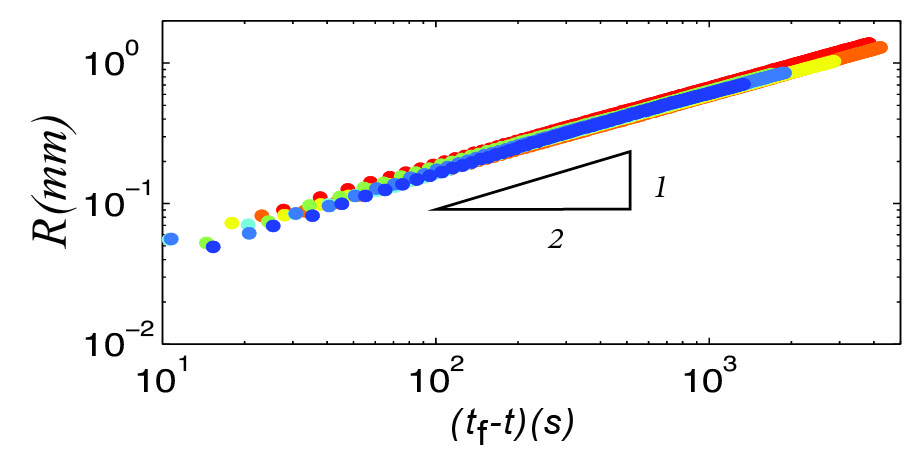}
\caption{The radius of the droplet plotted against $t_f-t$ with $t_f$ the lifetime of the droplet and $t$ the actual time. The triangle indicates the $1/2$-power law, the dots represent the data sets for 7 different experiments, where the number of particles was varied. For a certain $\hat{t}<t_f$ the final ball size $R_{ball}$ is reached. The final time was extrapolated as $t_f=\hat{t}+R_{ball}^2/D'$.}
\label{fig3}
\end{figure}

To understand the final structure of these supraballs, it will turn out crucial to understand the dynamics of the droplet evaporation. The fact that we do not observe shell formation, suggests that the particles do not influence the droplet evaporation. To test this hypothesis, we compare the evaporation dynamics to that of a liquid drop that does not contain any particles. The evaporative mass loss from such a drop is typically governed by the diffusion of vapor molecules in the surrounding air \cite{Deegan:1997, Popov:2005, gelderblom2011water}.
For diffusion-limited evaporation, the rate of volume change of the drop is given by
\begin{equation}
\frac{dV}{dt}\sim D' R,\label{vr}
\end{equation}
where $R$ is the drop radius, and $D'=D_{va} \Delta c /\rho$, with $D_{va}$ the diffusion constant for vapor in air, $\Delta c$ the vapor concentration difference between drop surface and the surroundings and $\rho$ the liquid density \cite{MarinPRL2011}.
One might have expected the evaporation rate from the drop surface to be proportional to the droplet surface area $\sim R^2$. However, the vapor concentration gradient is proportional to $1/R$, and therefore the total evaporation rate is proportional to $R$ \cite{Eggers:2010}.  If the droplet evaporates with a constant contact angle, we find that, since $V\sim R^3$,

\begin{equation}
R(t) \sim [D'(t_f-t)]^{1/2}\label{eqR}.
\end{equation}
Here $t_f$ is the total droplet lifetime in case no particles are present, for which the drop radius reaches zero. In the present case the drop radius saturates at a finite radius, $R_{ball}$, at a time $\hat{t}=t_f - R_{ball}^2/D'$, corresponding to the moment where the particles become densely packed. In figure \ref{fig3} we plot the colloidal droplet radius versus $t_f -t$. Our experimental data for different number of particles are in very good agreement with the $1/2$-power law. This confirms that the particles do not influence the evaporation process, until the final radius $R_{ball}$ is reached. The scaling (\ref{vr}) implies that the speed with which the interface is moving inwards, is given by $dR/dt\sim D'/R$.
Hence, the interface speed increases dramatically as the droplet shrinks and the maximum speed reached in the experiment will be determined by the final radius $R_{ball}$. As we will show further on, this increase in interface speed determines the particle packing inside the supraballs.

\begin{figure}[h*]
	\includegraphics[width=3.3 in]{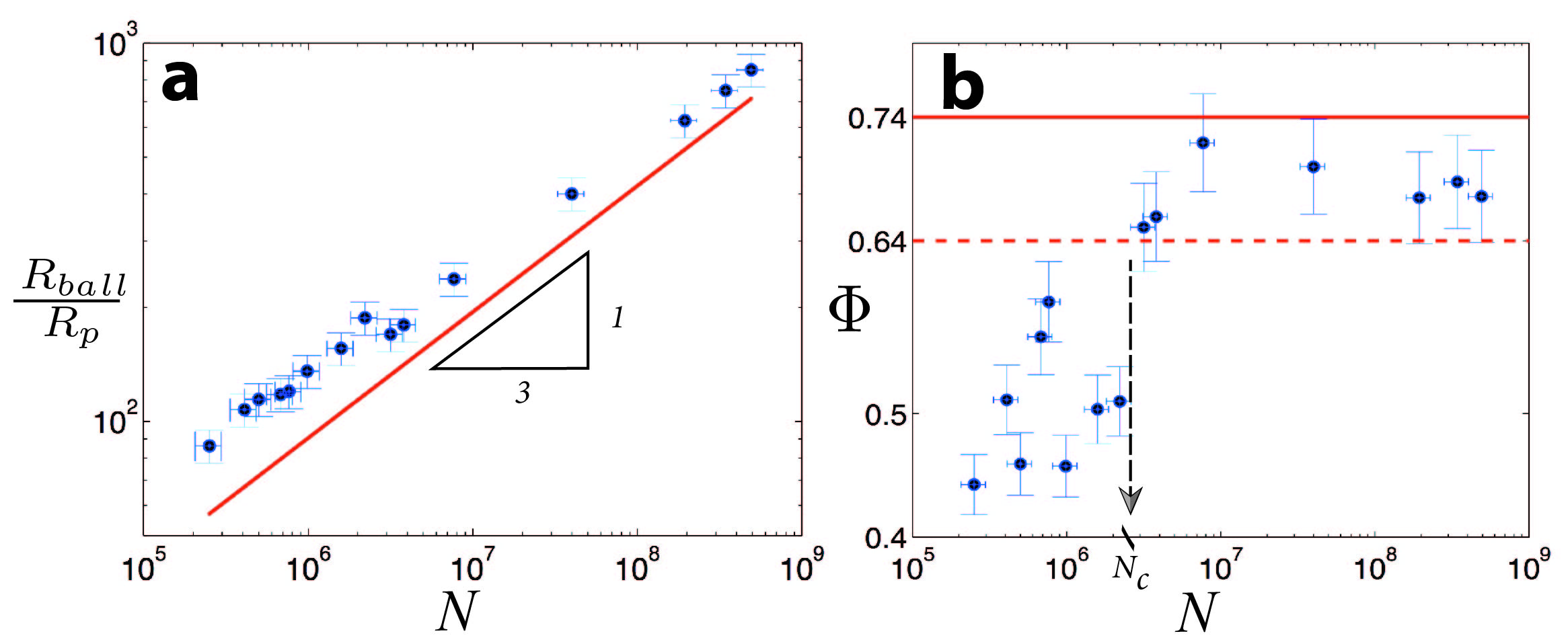}
\caption{ (a) Supraball to microparticle diameter $R_{ball}/R_{p}$ plotted against the total amount of particles $N$ in the system. (b) Packing fraction $\Phi$ strongly depends on $N$. Blue dots represent experimental measurements and the red solid line corresponds the most efficient particle packing $\Phi=0.74$ (hexagonal close packed), the dashed line marks $\Phi=0.64$, random close packed. $N_c$ is the critical number of particles, above which we find an ordered ball structure}
\label{fig4}
\end{figure}

The final size of the ball depends on the number of particles inside the drop. This can be tuned by manipulating either the initial particle concentration or the droplet size. In our experiment, the ball size was in the range  $100<R_{ball}/R_{p}<1000$, with $R_p$ the particle radius. Clearly, the exact final size of the ball will not only depend on the amount of particles in the system but also on their packing fraction.
We define the packing fraction as
\begin{equation}
\Phi\equiv N \left(\frac{R_p}{R_{ball}}\right)^3,
\end{equation}
where $N$ is the total number of particles in the droplet. The final supraball radius $R_{ball}$ is accurately determined from Scanning Electron Microscope (SEM) images. If the packing fractions were identical for all supraballs, one would expect that $R_{ball}/R_p\sim N^{1/3}$, as depicted in figure \ref{fig4}a \footnote{All quantitative results shown in this  paper have been performed with colloids of $1\mu m$ diameter (non-surface-modified fluorescent microspheres supplied by Thermo Scientific), but the same qualitative behavior has been observed for $0.2$ and $2\mu m$. The colloidal solutions were always prepared with deionized water.}. However, in figure \ref{fig4}b we see that the packing fraction strongly depends on the number of particles in the system.

As the number of particles increases, the packing fraction approaches that of a perfect Hexagonal Close Packing configuration, in which case one would find $\Phi=0.74$ \cite{Kepler1611}, hence, we have an \emph{ordered} particle packing inside the balls. On the other hand, the supraballs with a smaller amount of particles show remarkably low packing fractions, even below the Random Close Packing (RCP) limit ($\Phi=0.64$) \cite{weitzpacking}, corresponding to a \emph{disordered} particle arrangement. The balls which show packing fractions below the RCP limit contain several  empty cavities.
Remarkably, the final configuration reached, seems to depend on the number of particles in the system. In figure \ref{fig4}b we indicated the critical number of particles $N_c \approx 3\cdot 10^6$ when the packing fraction reaches that of a RCP. For $N<N_c$ we get a loose, \emph{disordered} particle packing in the supraball, whereas for $N>N_c$ we get a densely packed, \emph{ordered} supraball.

What causes the transition from ordered to disordered packings, and what determines the critical number of particles? To explain this, we follow a similar approach as in Mar\'in et al. \cite{MarinPRL2011}: we compare the time scale on which particles can arrange by diffusion to the hydrodynamic time scale for the particle transport by convection, given by the inward motion of the liquid-air interface.
If the diffusion time is small compared to the hydrodynamic time, particles can arrange into an ordered packing. The diffusive time scale is $t_d=R_p^2/D$, with $R_p$ the particle radius and $D$ the diffusivity of the particles in the liquid \cite{MarinPRL2011}. The hydrodynamic time-scale is  $t_h=L/\left|\frac{dR}{dt}\right|$. Here $R(t)$ is the droplet radius, and $L$ is the typical inter-particle distance, which depends on the particle concentration as $L =N^{-1/3} R$, as long as the solution is dilute ($L\gg R_p$). We define the ratio of both time scales as:
\begin{equation}
\mathcal{A}(t)\equiv\frac{t_d}{t_h}=\left|\frac{dR(t)}{dt}\right| \frac{t_d}{L}=\frac{D'}{D} N^{1/3} \left(\frac{R_p}{R(t)}\right)^2,\label{A}
\end{equation}
where in the last step we used (\ref{vr}) to replace $dR/dt \sim D'/R$.

From (\ref{A}) we observe that $\mathcal{A}(t)$ increases as the droplet radius becomes smaller during the evaporation (see figure \ref{fig3}), until the limit $R=R_{ball}$ is reached.  A cross-over between the time-scales is reached when the hydrodynamic time becomes equal to the diffusion time, hence when $\mathcal{A}=1$. If the cross-over is reached when $R\gg R_{ball}$, the amount of crystalline clusters is still very small. From this point in time onwards the interface speed is too high for the particles to further arrange in a crystalline way \cite{MarinPRL2011}. Instead, they are pushed together in a random arrangement, with a low packing fraction. If the cross-over is reached when $R\le R_{ball}$, the particle packing is already dense and ordered, and we find a high packing fraction. For all droplets the evaporative mass loss, and hence the decrease in radius, is the same (see figure \ref{fig3}), hence, the moment when the particle packing becomes sufficiently dense for particles to arrange depends solely on the number of particles in the droplet. If $N$ is high ($N>N_c$), this moment is reached relatively early, i.e. well before $\mathcal{A}=1$, and we get an ordered particle packing inside the supraballs.
Using that $R_{ball}/R_p\sim N^{1/3}$ and considering $\mathcal{A}=1$, we find from (\ref{A}) the critical number of particles above which we obtain ordered supraballs

\begin{equation}
N_c\sim\left(\frac{D'}{D}\right)^3.
\end{equation}
This result emphasizes that the transition is governed by two diffusion processes: the diffusion of vapor, determining the speed of evaporation, versus the diffusion of particles inside the drop. The ratio of diffusion constants selects the critical number of particles. In our experiment $D'=3\times 10^{-10}$ m$^2$/s and $D=2\times 10^{-13}$ m$^2$/s, from which we find that $N_c\sim 10^9$. This is 2 to 3 orders of magnitudes larger than the experimentally observed $N_c$. However, in the preceding analysis we have neglected all prefactors, and we emphasize that the result is strongly (to the third power) dependent on the experimental parameters included in $D'$, i.e. humidity, liquid density, diffusivity of vapor, and saturated vapor concentration.

\begin{figure}[h]
	\includegraphics[width=3.4 in]{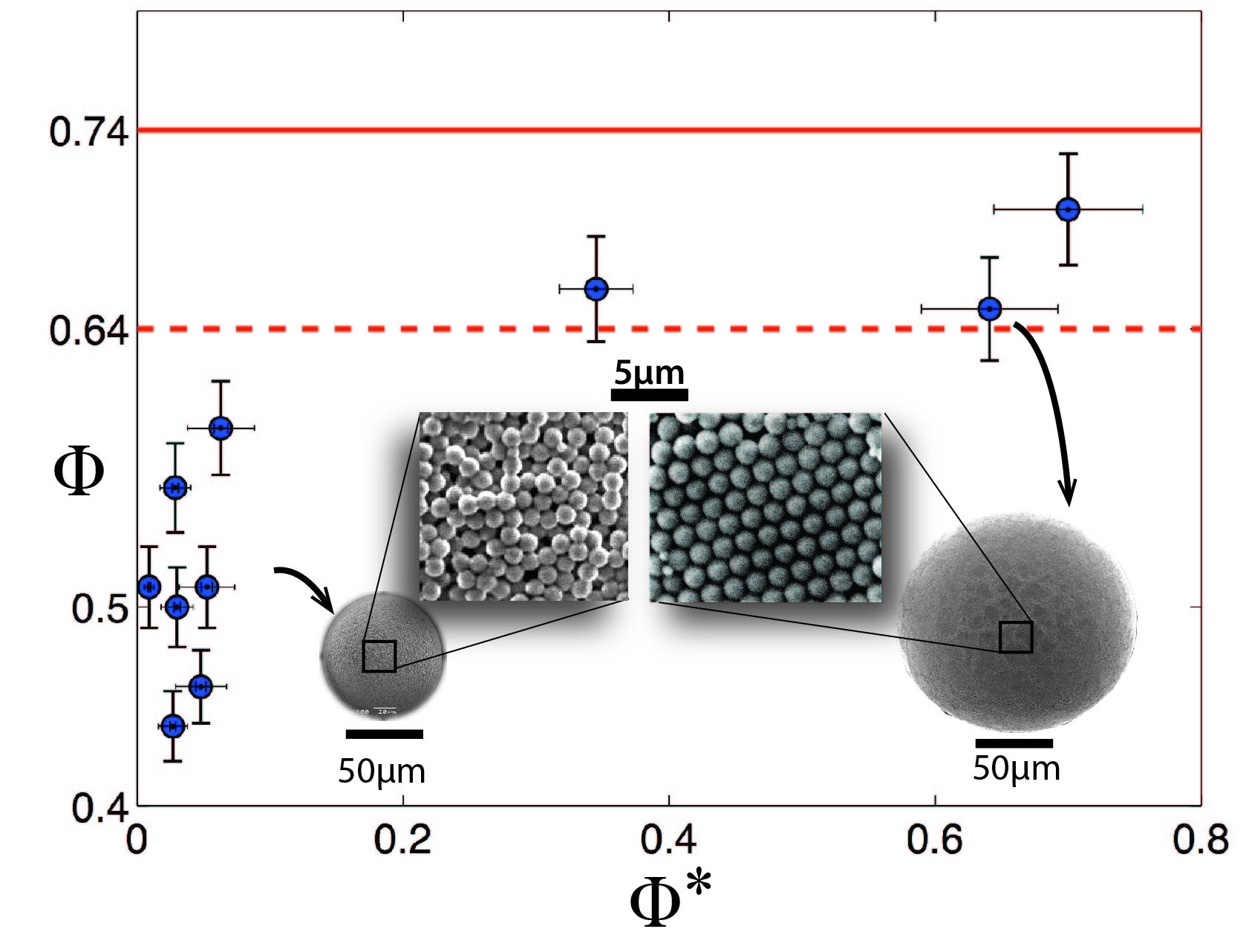}
\caption{Final packing fraction $\Phi_f$ versus the packing fraction at the cross-over time $\Phi^*$.  Droplets below a certain $\Phi^*$ have a too low packing fraction at the cross-over time to achieve final packing fractions above the RCP limit. The particle packing can not only be obtained from the value of the packing fraction, but it can also directly be observed from the SEM images of the surface of the supraballs.}
\label{fig5}
\end{figure}

To verify whether the final packing fraction indeed depends on the spacing between the particles the moment the cross-over time is reached, we go back to our experimental data. We define the time-dependent packing fraction as $N(R_p/R(t))^3$.
As the droplet evaporates, this packing fraction will increase until it reaches its final value $\Phi$. At the cross-over, defined by ${\cal A}=1$, the droplets will have a packing fraction $\Phi^*$. If this $\Phi^*$ is low, the amount of crystalline clusters is still very small. On the other hand, if $\Phi^*$ is high, we expect crystalline clusters to have formed already. After the cross-over time, the interface moves too fast to allow for further ordering, and it just presses the ordered particle clusters closer together.
In figure \ref{fig5} we show that droplets with a high $\Phi^*$ have a high $\Phi$: when $\Phi^* \gtrapprox 0.1$ we obtain a final packing fraction above the RCP limit.

We cannot predict the critical $\Phi^*$ theoretically. However, we can, retrospectively, use the experimental critical $\Phi^*$ to compute $N_c$. Using that $R_{ball}/R_p\sim 0.1 N^{1/3}$ at the cross-over,
we obtain $N_c\sim 10^7$, which is in the same order of magnitude as our experimental results; see figure \ref{fig4}b.

The particle packing in the supraballs can not only be assessed by measuring the packing fraction, but it can also directly be seen in SEM images from the surface of the colloidal supraballs, as shown in figure \ref{fig5}. The size of the soccer-ball-like crystalline patches on the surface of the ordered supraballs depends on the ball size: bigger balls will show larger patches due to the reduced curvature at their surfaces.
To explain the size of the crystalline domains, we hypothesize that a crystalline patch will bend radially no more than a particle size. Then, it follows by simple trigonometry that the size of a patch $S$ will be related to the ball radius $R_{ball}$ and the particle size $R_p$ via: $S/R_{ball}=\arccos(1-R_p/R_{ball})$. This expression predicts a typical patch size of $\sim15\mu m$ for a ball with $R_{ball}=100 \mu m$ and $R_p=1\mu m$, which is in the right order of magnitude as one can observe in figure \ref{fig2}d.

In conclusion, in this Letter we devise a simple technique to create spherical colloidal supraballs relying only on droplet evaporation over a robust superhydrophobic surface. The supraballs show a highly ordered structure if the number of particles inside the drop is large enough to trigger early particle clustering.
The critical number of particles required to obtain an ordered particle packing inside the balls depends on the parameters driving the droplet evaporation (through $D'$)  and the diffusivity of the particles. Hence, by controlling the humidity and ambient temperature the supraball packing fraction and hence size can be controlled. Massive fabrication of micro-compact-supraballs could easily be achieved by simply spraying a colloidal solution over the micro-structure in a controlled atmosphere. By tuning the wetting properties of the particles one could also be able to generate the well-known \emph{colloidosomes} \cite{colloidosomes} using the same proposed technique.

\begin{acknowledgements}
We gratefully acknowledge Vicenzo Vitelli, Martin van Hecke, Devaraj van der Meer, Chao Sun and many others for their useful comments and their support. The authors also acknowledge financial support by the NWO grant No. 700.10.408 and No. 700.58.041.
\end{acknowledgements}



\bibliography{suprabib.bib}
\end{document}